\DocumentMetadata{
 pdfversion=2.0,
 lang = en
 }
\documentclass[sigconf,nonacm]{acmart}
\AtBeginDocument{%
  }

\usepackage{tabularx}
\usepackage{multirow} 
\usepackage[table]{xcolor}
\newtheorem{definition}{Definition}[section]

\usepackage{cleveref}
\usepackage{enumitem}

\usepackage{caption}
\usepackage{subcaption}
\usepackage{booktabs} 

\usepackage{listings}
\definecolor{eclipseStrings}{RGB}{42,0.0,255}
\definecolor{eclipseKeywords}{RGB}{127,0,85}
\colorlet{numb}{magenta!60!black}
\lstdefinelanguage{json}{
    basicstyle=\scriptsize\ttfamily,
    commentstyle=\color{eclipseStrings}, % style of comment
    stringstyle=\color{eclipseKeywords}, % style of strings
    numberstyle=\scriptsize,
    stepnumber=1,
    numbersep=8pt,
    showstringspaces=false,
    breaklines=true,
    frame=lines,
    backgroundcolor=\color{white}, %only if you like
    string=[s]{"}{"},
    comment=[l]{:\ "},
    morecomment=[l]{:"},
    escapechar=|,
    literate=
        *{0}{{{\color{numb}0}}}{1}
         {1}{{{\color{numb}1}}}{1}
         {2}{{{\color{numb}2}}}{1}
         {3}{{{\color{numb}3}}}{1}
         {4}{{{\color{numb}4}}}{1}
         {5}{{{\color{numb}5}}}{1}
         {6}{{{\color{numb}6}}}{1}
         {7}{{{\color{numb}7}}}{1}
         {8}{{{\color{numb}8}}}{1}
         {9}{{{\color{numb}9}}}{1}
}

\usepackage{tikz}
\usetikzlibrary{arrows.meta}

\begin{document}

\title{ProvCards: Artifact-Oriented Provenance for Heterogeneous Scientific Workflows}

\author{Tanja Auge}
\orcid{0009-0006-2150-9713}
\email{tanja.auge@ur.de}
\affiliation{%
  \institution{University of Regensburg}
  \country{Germany}
}

\author{Marinus Holitschke}
\orcid{0009-0004-6721-8108}
\email{marinus.holitschke@stud.uni-regensburg.de}
\affiliation{%
  \institution{University of Regensburg}
  \country{Germany}
}

\author{Meike Klettke}
\orcid{0000-0003-0551-8389}
\email{meike.klettke@ur.de}
\affiliation{%
  \institution{University of Regensburg}
  \country{Germany}
}

\renewcommand{\shortauthors}{Auge et al.}

\begin{abstract}
Reproducibility in scientific workflows is challenged by heterogeneous workflow components, execution environments, and fragmented provenance capture. Existing approaches provide established semantics and workflow-specific solutions, but provenance information often remains difficult to capture, exchange, and compose across workflow boundaries. This paper introduces \textit{ProvCards}, a novel artifact-oriented approach that treats provenance as a first-class artifact in scientific data-processing workflows. ProvCards can be generated directly by individual workflow components and independently stored, exchanged, reused, and composed, while remaining semantically grounded in the W3C PROV Data Model. This decouples provenance capture from any particular workflow system or centralized execution context and enables provenance to evolve incrementally alongside the artifacts it describes. Collections of ProvCards, called \textit{ProvDecks}, establish portable provenance scopes by composing related cards without merging or sacrificing their individual identities. \textit{ProvGraphs} provide derived representations for provenance reconstruction, querying, visualization, and analysis. Together, these representations enable provenance captured across heterogeneous components to be integrated into coherent and task-specific views, thereby facilitating cross-system traceability, provenance reuse, and reproducibility. 
We implement the approach using a graph-based provenance infrastructure and evaluate it in a scientific data-processing workflow. The evaluation demonstrates the feasibility of capturing provenance independently at the level of individual workflow components, composing it across component boundaries, and reconstructing provenance structures for querying and analysis. The results establish ProvCards as a foundation for portable, composable, and incrementally extensible provenance management in heterogeneous scientific workflows.
\end{abstract}



%\includegraphics[width=\textwidth, alt={Enjoying the baseball game from the third-base seats. Ichiro Suzuki preparing to bat.}]{sampleteaser}
%\caption{Seattle Mariners at Spring Training, 2010.}
%\label{fig:teaser}


\maketitle

\section{Introduction}

Provenance is increasingly important for verifying data and reproducing computational results. Modern scientific data processing, however, commonly spans databases, files, scripts, notebooks, workflow engines, external services, and distributed execution environments~\cite{Goble2020FAIRWorkflows,Kauppinen2020WorkflowDataProducts}. Since no single system controls or observes the complete processing lifecycle, the provenance of a result is typically distributed across multiple technical and organizational boundaries.
The participating systems already generate provenance-relevant information by design. Logs, execution traces, catalogs, and metadata records contain input and output references, commands, parameters, timestamps, software versions, responsible actors, and execution status. Yet this evidence is represented in system-specific formats and remains tied to the components that generated it. It therefore cannot readily be exchanged or composed into consistent end-to-end provenance. Existing approaches provide standardized provenance semantics, workflow-integrated capture, provenance processing, or workflow-level packaging, but they lack an independently manageable unit for incrementally capturing and composing provenance across heterogeneous components.

We address this gap with the \textit{ProvCard Model}, an artifact-oriented approach grounded in the W3C PROV Data Model~\cite{provDM}. A \textit{ProvCard} is an independently manageable unit of provenance capture and exchange. Each card represents a provenance-relevant entity, activity, or agent and can refer explicitly to elements captured by other components. ProvCards can be generated incrementally during or after execution and can be identified, stored, retrieved, and exchanged independently of a shared workflow environment. Their contribution lies not in introducing new provenance semantics, but in providing a portable operational unit for provenance that remains usable when no single system controls the complete data-processing lifecycle.
Within this layered model, ProvCards form the atomic elements. Collections of related cards, called \textit{ProvDecks}, establish portable scopes for composing and exchanging provenance while preserving the identity of the individual cards. \textit{ProvGraphs} provide derived graph representations for reconstructing, querying, visualizing, and analyzing the relationships captured by the cards.

\begin{figure*}
    \centering
    \includegraphics[width=.975\linewidth]{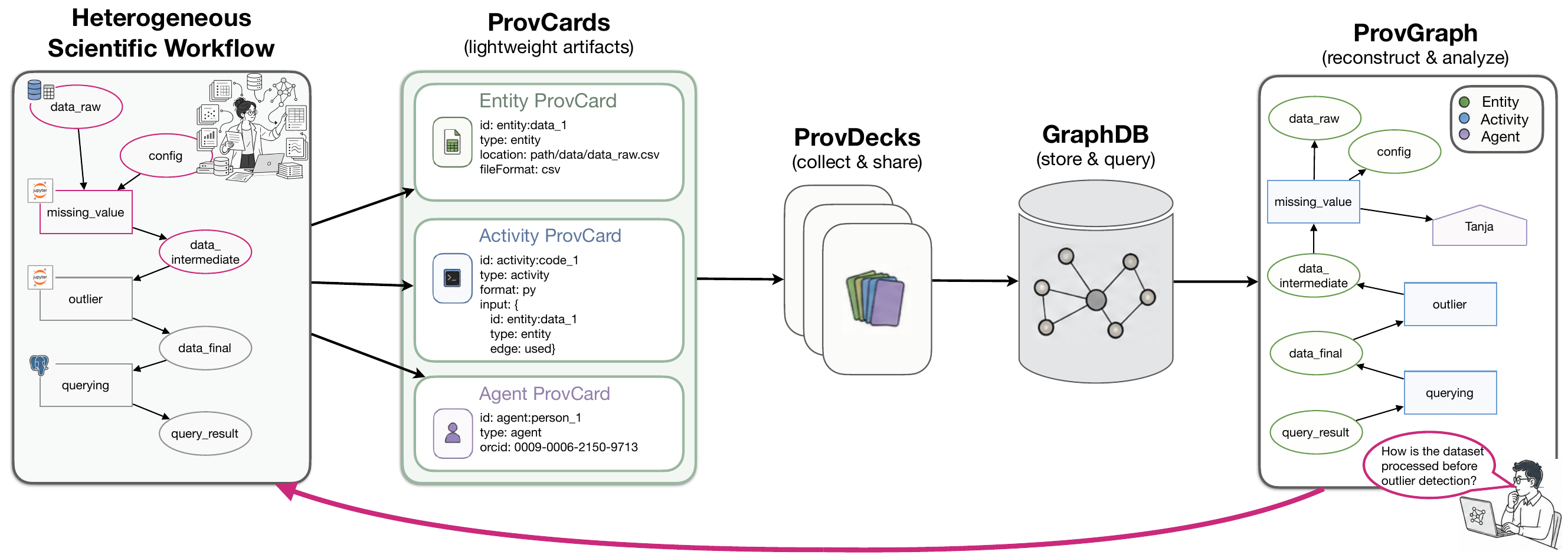}
    \caption{From components distributed across heterogeneous systems to composable provenance artifacts. Independent data-processing components generate ProvCards for the provenance elements they observe. Related cards can be collected in ProvDecks, stored, and reconstructed as a ProvGraph for querying and analysis. The highlighted elements illustrate an exemplary provenance question and the provenance path providing its answer.}
    \label{fig:teaser}
\end{figure*}

Figure~\ref{fig:teaser} illustrates the proposed approach. The left-hand side shows data artifacts and processing activities distributed across independently operating systems, including a database, a script or notebook environment, and an external service. Each component generates ProvCards for the provenance elements it observes. Related cards can subsequently be collected in a ProvDeck, stored, and reconstructed as a ProvGraph. The highlighted elements illustrate an exemplary provenance question: the relevant path through the reconstructed graph identifies the data and processing steps that contributed to a selected result. Fragmented execution evidence thus becomes connected and queryable without requiring all processing steps to be executed or governed by a single workflow system.

The main contributions of this paper are:
\begin{enumerate}
\item An \textbf{artifact-oriented provenance model} that defines an independently manageable unit of provenance capture and exchange, extending provenance management beyond the control boundaries of individual database and workflow systems (Section~\ref{sec:4-provModel}).
\item A \textbf{layered representation consisting of ProvCards, \\ ProvDecks, and ProvGraphs} for incremental capture, cross-system composition, controlled sharing, and analysis (Sections~\ref{sec:4-provModel:ProvenanceArtifacts} \& \ref{sec:4-provModel:ProvGraph}).
\item \textbf{Formalized transformations and information- \\ preser\-va\-tion properties} between the three representations, including information reduction through abstraction, anonymization, and selective disclosure (Section~\ref{sec:4-provModel:ProvenanceTramsformation}).
\item An \textbf{implementation and empirical validation} demonstrating independent provenance capture, composition, reconstruction, and querying in a scientific data-processing workflow (Sections~\ref{sec:5-implementation}--\ref{sec:7-validation}).
\end{enumerate}

\section{Related Work}
\label{sec:2-relatedWork}

To position the proposed artifact-oriented abstraction, we first examine how existing work addresses provenance semantics, capture, packaging, and lifecycle management. This comparison establishes the dimensions from which the challenges and requirements in Section~\ref{sec:3-challenges} are derived.

\paragraph{\textbf{Provenance Models and Capture.}}
The \textit{W3C PROV standard} provides a uniform and machine-interpretable model for representing provenance. PROV-DM~\cite{provDM} distinguishes \emph{entities}, \emph{activities}, and \emph{agents}, with relationships such as \texttt{used}, \texttt{wasGeneratedBy}, and \texttt{wasDerivedFrom} capturing dependencies between them. PROV-O~\cite{provo} provides an OWL-based representation of these concepts. This family of standards establishes interoperable provenance semantics and serialization mechanisms, but they do not prescribe how provenance fragments generated independently by heterogeneous components should be captured, managed, and composed throughout their lifecycle.

Beyond such workflow-integrated capture, workflow engines, applications, operating systems, and supporting infrastructure generate logs and execution traces containing timestamps, commands, parameters, software versions, resource references, and execution status. These records constitute valuable provenance evidence even when provenance capture was not an explicit objective of the executing component. Their formats, identifiers, granularity, and semantics nevertheless differ across systems. Consequently, turning these records into consistent end-to-end provenance requires component-specific extraction and mapping.

\paragraph{\textbf{Provenance Processing and Packaging.}}
Provenance-processing frameworks operate on structured provenance representations. ProvToolbox~\cite{provtoolbox}, for example, supports the creation, validation, and transformation of PROV documents, while PROV-Template~\cite{prov-template} enables declarative generation of provenance graphs. Such frameworks provide important processing capabilities but assume that the provenance information to be processed has already been mapped to an appropriate representation. Their primary purpose is therefore not to define a lifecycle for independently generated provenance fragments.

These approaches demonstrate that workflow provenance can be represented and exchanged independently of the technology used to execute a workflow. Their primary aggregation scope, however, is typically a crate, research object, or workflow execution. This scope may not yet be available when an individual component generates provenance evidence. This leaves open how independently generated provenance fragments can be managed before a complete workflow-level object is available and how they can later be incorporated into such an object.

\paragraph{\textbf{Documentation.}}
Documentation artifacts such as \textit{Datasheets for Datasets}~\cite{modelCards-1}, \textit{Model Cards}~\cite{ModelCards-2}, \textit{Data Cards}~\cite{dataCards-1}, and related dataset-documentation initiatives~\cite{dataCards-2} improve the transparency and reuse of individual research assets. They describe properties such as an artifact’s intended use, composition, creation process, evaluation, limitations, licensing, and ethical considerations. Such documentation is related to provenance because it records contextual information about the origin and use of an artifact. Its primary focus, however, is the description of a dataset or model as a static resource. It does not generally represent the dynamic relationships among data, activities, agents, intermediate artifacts, and execution events that arise while a workflow is running. Documentation artifacts and dynamic provenance representations therefore address complementary concerns: the former communicate the context, intended use, and limitations of a resource, whereas the latter represent relationships among artifacts, processing activities, and responsible agents.

\paragraph{\textbf{Controlled Disclosure.}}
Provenance may contain sensitive information about datasets, users, workflow logic, and execution environments. Prior work distinguishes data, module, and structural privacy in scientific workflow provenance \cite{provenancePrivacy}. This motivates a distinction between provenance capture and disclosure: information may need to be omitted, abstracted, or anonymized before provenance is exchanged.

\paragraph{\textbf{Research Gap.}}
Existing approaches provide standardized provenance semantics, workflow-integrated capture, provenance processing, documentation, and workflow-level packaging. However, provenance evidence generated by independently operating components remains coupled to system-specific records or is managed only after a workflow-level aggregation scope has been established. The approaches discussed above do not make an independently generated provenance fragment the primary unit of lifecycle management and subsequent cross-system composition. Consequently, it remains difficult to persistently identify, store, exchange, and compose such fragments across heterogeneous system boundaries.
To address this gap, the ProvCard Model introduces an independently manageable, artifact-level unit for provenance fragments grounded in PROV-DM. Its contribution concerns the granularity and lifecycle of provenance management rather than new provenance semantics. ProvCards are intended to complement, rather than replace, workflow-integrated capture, PROV documents, and workflow-level packages such as Research Objects. The resulting challenges and requirements are derived in Section~\ref{sec:3-challenges}; the model itself is introduced in Section~\ref{sec:4-provModel}.
\section{Challenges and Requirements}
\label{sec:3-challenges}

\paragraph{\textbf{Challenges.}}
Existing approaches leave five challenges for provenance management in heterogeneous scientific workflows.

\newcounter{challenge}
\renewcommand{\thechallenge}{C\arabic{challenge}}

\begin{itemize}[leftmargin=3.75em]
\refstepcounter{challenge}
\item[\textbf{\thechallenge}]\label{item:C1}
\textbf{Heterogeneous workflow environments.}
Scientific workflows combine workflow engines, scripts, databases, external services, and distributed infrastructures whose provenance mechanisms are often incompatible.

\refstepcounter{challenge}
\item[\textbf{\thechallenge}]\label{item:C2}
\textbf{Independent and fragmented provenance capture.}
Workflow components capture provenance independently, using different formats and levels of detail, which complicates composition and end-to-end reconstruction.

\refstepcounter{challenge}
\item[\textbf{\thechallenge}]\label{item:C3}
\textbf{Limited portability and composability of provenance.}
Although PROV-DM provides interoperable semantics, provenance is often embedded in workflow-specific records or packages and cannot be exchanged and composed independently.

\refstepcounter{challenge}
\item[\textbf{\thechallenge}]\label{item:C4}
\textbf{Missing lifecycle for provenance artifacts.}
Independently generated provenance fragments lack explicit mechanisms for persistent identification, storage, retrieval, exchange, and composition.

\refstepcounter{challenge}
\item[\textbf{\thechallenge}]\label{item:C5}
\textbf{Provenance detail vs.\,capture overhead.}
Detailed provenance improves reconstruction and analysis but increases computational and organizational overhead. Capture must therefore remain lightweight for integration into existing workflows.
\end{itemize}

\paragraph{\textbf{Requirements.}}
Based on these challenges, we derive seven requirements for artifact-oriented provenance management across heterogeneous scientific workflows: 

\newcounter{requirement}
\renewcommand{\therequirement}{R\arabic{requirement}}

\begin{itemize}[leftmargin=3.75em]
\refstepcounter{requirement}
\item[\textbf{\therequirement}]\label{item:R1}
\textbf{Machine-readable and Extensible Representation.}
Represent PROV-DM entities, activities, agents, and relationships in a structured, machine-readable format extensible with domain-specific metadata.
\refstepcounter{requirement}
\item[\textbf{\therequirement}]\label{item:R2}
\textbf{Engine Independence.}
Operate independently of workflow engines, programming languages, and software environments without requiring replacement of existing infrastructure.
\refstepcounter{requirement}
\item[\textbf{\therequirement}]\label{item:R3}
\textbf{Lightweight and Incremental Capture.}
Allow workflow components to generate provenance artifacts independently and with limited overhead.
\refstepcounter{requirement}
\item[\textbf{\therequirement}]\label{item:R4}
\textbf{Support for Heterogeneous Resources.}
Provide a common abstraction for datasets, scripts, databases, computational models, and external services.
\refstepcounter{requirement}
\item[\textbf{\therequirement}]\label{item:R5}
\textbf{Provenance Artifact Management.}
Support persistent identification, storage, retrieval, and exchange independently of the generating environment.
\refstepcounter{requirement}
\item[\textbf{\therequirement}]\label{item:R6}
\textbf{Portability and Composability.}
Preserve artifact identities and relationships during exchange and composition across systems and institutions.
\refstepcounter{requirement}
\item[\textbf{\therequirement}]\label{item:R7}
\textbf{Provenance Reconstruction and Analysis.}
Support reconstruction of provenance graphs for dependency, lineage, and responsibility analysis.
\end{itemize}

\section{The Provenance Model}
\label{sec:4-provModel}

We address the requirements identified in Section~\ref{sec:3-challenges} through an artifact-oriented provenance model to independently capture, exchange, and compose provenance in heterogeneous workflow environments.

\begin{figure*}[ht]
\centering
\scalebox{.55}{
\begin{tikzpicture}[
    >=Stealth,
    font=\small,
    activity/.style={
        rectangle,
        rounded corners=2pt,
        draw=blue!70!black,
        fill=blue!10,
        minimum width=3.2cm,
        minimum height=1.2cm,
        align=center
    },
    entity/.style={
        rectangle,
        draw=black,
        minimum width=3.2cm,
        minimum height=1.0cm,
        align=center
    },
    arrow/.style={
        ->,
        thick
    }
]

% Entitäten
\node[entity] (raw) at (0,  1.3) {\texttt{data\_raw.csv}};
\node[entity] (config) at (0, -1.3) {\texttt{config.txt}};

% Aktivitäten
\node[activity] (missing) at (4.5, 0) {\texttt{missing\_values.py}};
\node[entity]   (intermediate) at (9, 0) {\texttt{data\_intermediate.csv}};
\node[activity] (outlier) at (13.5, 0) {\texttt{outlier.py}};
\node[entity]   (final) at (18, 0) {\texttt{data\_final.csv}};

% Eingänge in missing_values.py
\draw[arrow] (raw.east) -- (missing.north west);
\draw[arrow] (config.east) -- (missing.south west);

% Pipeline
\draw[arrow] (missing.east) -- (intermediate.west);
\draw[arrow] (intermediate.east) -- (outlier.west);
\draw[arrow] (outlier.east) -- (final.west);
\end{tikzpicture}}
\caption{Overview of the data preprocessing workflow used in the case study.}
\label{fig:workflow}
\end{figure*}
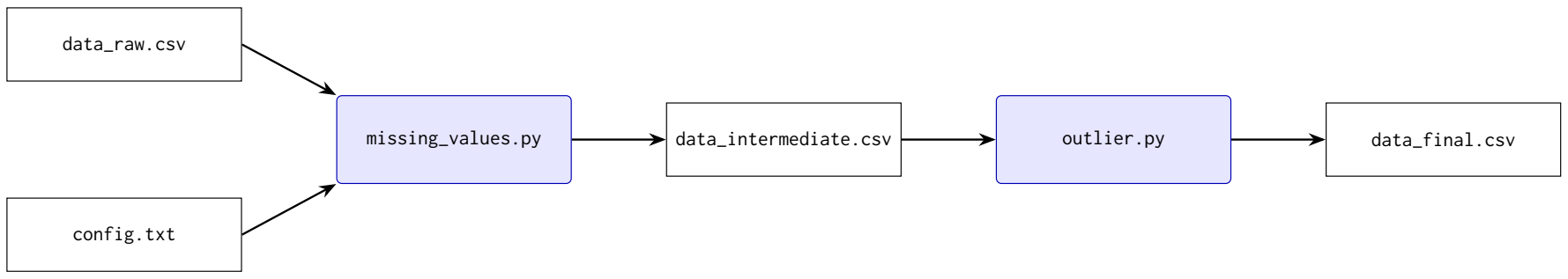

%%%%%%%%%%%%%%%%%%%%%%%%%%%%%%%%%%%%%%%%%%%%%%%%%%%%%%%%%%%%%%

\paragraph{\textbf{Design Principles.}}
The design of the proposed provenance model follows five principles that support provenance capture and exchange across heterogeneous workflow environments:
\begin{itemize}
\item \textit{Independence} enables individual workflow components to generate provenance artifacts without requiring access to a global workflow representation or centralized provenance service.
\item \textit{Portability} ensures that provenance artifacts can be stored and exchanged independently of the workflow system that generated them.
\item \textit{Composability} enables independently generated provenance artifacts to be combined into larger provenance structures while preserving the identity and semantics of the individual artifacts.
\item \textit{Immutability} ensures that a provenance artifact represents a specific observation of a workflow component at a particular point in time. Changes to the underlying workflow, data, or metadata therefore result in new provenance artifacts rather than modifying existing ones.
\item \textit{Lightweightness} allows provenance to be captured incrementally by individual workflow components without requiring a complete workflow representation or centralized provenance infrastructure.
\end{itemize}

%%%%%%%%%%%%%%%%%%%%%%%%%%%%%%%%%%%%%%%%%%%%%%%%%%%%%%%%%%%%%%
\subsection{Provenance Artifacts}
\label{sec:4-provModel:ProvenanceArtifacts}

The ProvCard Model introduces a layered, artifact-oriented representation for provenance management across heterogeneous scientific workflows. It preserves the provenance semantics of PROV-DM~\cite{provDM}, while introducing an explicit artifact-level abstraction for provenance capture, composition, exchange, and analysis. The model distinguishes three complementary representations: \textit{ProvCards}, \textit{ProvDecks}, and \textit{ProvGraphs}. ProvCards represent independently manageable provenance artifacts, ProvDecks provide a composition and exchange scope for collections of artifacts, and ProvGraphs provide a derived representation for provenance analysis.

\paragraph{\textbf{ProvCards.}}
A \textit{ProvCard} represents the fundamental provenance artifact of the proposed model. Each ProvCard describes a single provenance-relevant element corresponding to an entity, activity, or agent according to PROV-DM.
ProvCards are designed as independent and self-contained artifacts. 
For example, a measurement instrument may generate a ProvCard describing a raw dataset, while a processing tool creates another ProvCard describing a transformation activity. 
The relationship between both artifacts is established through explicit provenance references rather than through a shared workflow environment.

Let $\mathcal{T}$ denote the set of PROV types: $\mathcal{T}=\{\texttt{entity},\texttt{activity},$ $\texttt{agent}\}$. Let $\mathcal{T}_S$ denote the set of Prov subtypes, i.e., $\mathcal{T}_S=$ \{\texttt{data}, \texttt{metadata}, \texttt{policy}, \texttt{code}, \texttt{institution}, \texttt{person}, ...\}, where the first three labels define entities, and the last two labels define agents. \texttt{code} is a specialized activity. The available subtypes depend on the underlying ontology. Let further $\mathcal{E}$ denote the set of supported provenance relation types based on PROV-DM: $\mathcal{E}=$\{\texttt{used}, \texttt{was-} \texttt{GeneratedBy}, \texttt{wasAssociatedWith}, \texttt{wasAttributedTo}, \texttt{actedOn-} \texttt{BehalfOf}, \texttt{was-} \texttt{DerivedFrom}, \texttt{wasInformedBy}\}.

\begin{definition}[ProvCard]
Let $\mathcal{T}$ denote the set of PROV types and $\mathcal{E}$ the set of PROV relation types.
Let $\mathcal{K}_C$ denote a set of keys and $\mathcal{V}_C$ a set of atomic values.
Let $id_C$ be a unique identifier, $t_C \in \mathcal{T}$ a PROV type, $t_S \in \mathcal{T}_S$ a PROV subtype, and $ A_C=\left\{(a_i,v_i) \;\middle|\; a_i\in\mathcal{K}_C,\right.$ $\left.v_i\in\mathcal{V}_C\right\}$, a set of key-value pairs.
Let further $R_{\mathrm{in}}$ and $R_{\mathrm{out}}$ be sets of relationships $r_i$, where each relationship is defined as $r_i=(id,t,e_i)$, with $id$ a unique identifier, $t\in\mathcal{T}_S$, and $e_i\in\mathcal{E}$.
Then, a \textit{ProvCard} is defined as the tuple $C=(id_C,t_C,t_S,A_C,R_{\mathrm{in}},R_{\mathrm{out}})$.
\end{definition}

The set of key-value pairs $A$ consists of mandatory, conditionally mandatory, and optional provenance metadata. Mandatory attributes are \texttt{id}, \texttt{location}, \texttt{name}, and \texttt{type}. Conditionally mandatory attributes depend on the type of ProvCard. For entities, for example, \texttt{fileFormat} and \texttt{fileSize} are also mandatory. In contrast, attributes such as \texttt{runID} or \texttt{comment} are optional. Attributes describing directly connected ProvCards are likewise always mandatory.
The formal definition separates the local description of a provenance artifact from the relationships connecting it to other artifacts. In particular, $R_{\mathrm{in}}$ and $R_{\mathrm{out}}$ allow a ProvCard to refer to provenance information that is represented by independently created ProvCards. The resulting model therefore does not require all provenance information to be captured within a single workflow execution or provenance record.

In addition to general provenance metadata, domain-specific characteristics or excerpts from the data may be represented as \textit{quality information}. Such information extends the ProvCard without changing its underlying PROV semantics.

\begin{definition}[Quality ProvCard]
Let $C=(id_C,t_C,t_S,A_C,R_{\mathrm{in}},$ $R_{\mathrm{out}})$ be a ProvCard. Let $Q$ denote a set of quality-aspect identifiers, $\mathcal{K}_Q$ a set of quality attribute keys, and $\mathcal{V}_Q$ denote a set of atomic quality values. Let the set of structured quality values $\mathcal{S}$ be defined %recursively 
as $\mathcal{S} = \left\{ \{(k_1,v_1),\ldots,(k_n,v_n)\} \;\middle|\; n\geq 1,\; k_i\in\mathcal{K}_Q,\; v_i\in\mathcal{V}_Q \right\}$.
Let $A_Q = \left\{(q_i,w_i) \;\middle|\; q_i \in Q, w_i\in\mathcal{S}\right\}$, denote a finite set of quality aspects associated with the ProvCard.
Then, a \textit{quality ProvCard} is defined as the tuple $C_Q=(id_C,t_C,t_S,A_C,R_{\mathrm{in}},R_{\mathrm{out}},A_Q)$.
\end{definition}

Figure~\ref{fig:workflow} illustrates how independently generated provenance artifacts can describe the components of a workflow. The workflow consists of two activities (\texttt{missing\_values.py}, \texttt{outlier.py}), three datasets (\texttt{data\_raw.csv}, \texttt{data\_intermediate.csv}, \texttt{data\_final.} \texttt{csv}), and one metadata set (\texttt{config.txt}). Each component can be represented by a separate ProvCard, while explicit relationships connect the cards to form the provenance structure. An excerpt of the ProvCards is shown in Table~\ref{tab:example}. Note that $C_{\text{dataset}_2}$ is a quality ProvCard.

\paragraph{\textbf{ProvDecks.}}
Independently generated ProvCards can describe individual components of a workflow, but provenance exchange and analysis often require a defined scope containing multiple artifacts. To support this use case, related ProvCards can be combined into collections called \textit{ProvDecks}.
A ProvDeck represents a logical provenance scope for a set of artifacts associated with a common context, such as a workflow execution, experiment, dataset release, or provenance exchange scenario. It provides a composition and exchange boundary without merging the individual ProvCards into a single provenance record. The cards retain their individual identities and semantics and can therefore remain independently reusable outside the ProvDeck.
In this paper, composition denotes the collection of mutually compatible ProvCards into a common ProvDeck while preserving their identities and explicit provenance relationships. The model assumes that cards referring to the same provenance element use compatible identifiers and do not provide conflicting descriptions. Entity reconciliation and conflict resolution across independently administered identifier spaces are outside the scope of this work.

\begin{table*}[t]
\caption{Structure and attributes of the different ProvCard types.}
\label{tab:example}
\centering
\scalebox{.975}{$
\begin{array}{ll}
\multicolumn{1}{l}{\text{\textbf{ProvCards:}}} \\
C_{\text{dataset}_1} &=
(
\overbrace{\text{dataset}_1}^{=\,\textcolor{red}{id}},
\overbrace{\text{entity}}^{=\,\textcolor{red}{t}},
\overbrace{\text{data}}^{=\,\textcolor{red}{t_S}},
\overbrace{\{
(\text{name},\text{data\_raw.csv}),
(\text{location},\text{exp/data/data\_raw.csv})
\},}^{=\,\textcolor{red}{A_C}}
\overbrace{\emptyset}^{=\,\textcolor{red}{R_\text{in}}},
\overbrace{\emptyset}^{=\,\textcolor{red}{R_\text{out}}}
)
\\
C_{\text{metadata}_1} &=
(
\text{metadata}_1,
\text{entity},
\text{metadata},
\{
(\text{name},\text{config.txt}),
(\text{location},\text{exp/metadata/config.txt})
\},
\emptyset,
\emptyset
)
\\
C_{\text{activity}_1} &=
(
\text{activity}_1,
\text{activity},
\text{code},
\{
(\text{name},\text{missing\_values.py}),
(\text{location},\text{exp/activity/code/...})
\},
\{
(\text{dataset}_1,\text{data},\texttt{used}), \\
& \quad (\text{metadata}_1,\text{metadata},\texttt{used}), 
(\text{agent}_1,\text{agent},\texttt{wasAssociatedWith})
\},
\{
(\text{dataset}_2,\text{data},\texttt{wasGeneratedBy})
\}
)
\\
C_{\text{agent}_1} &=
(
\text{agent}_1,
\text{agent},
\text{person},
\{
(\text{name},\text{John Doe}),
(\text{location}, \text{company dataset})
\},
\emptyset,
\emptyset
)
\\
C_{\mathrm{dataset}_2} & = (
  \mathrm{dataset}_2,
  \mathrm{entity},
  \mathrm{data}, \{(\mathrm{name},\mathrm{data\_intermediate.csv}),
     (\mathrm{location},\mathrm{exp/data/data\_intermediate.csv})\},\\
& \quad \underbrace{\{(\mathrm{dataset1},\mathrm{entity}, \texttt{wasDerivedFrom})\}}_{=\textcolor{red}{R_\text{in}}}, \underbrace{\emptyset}_{=\textcolor{red}{R_\text{out}}}, \underbrace{\{
(\mathrm{gaps},
 \{
  (\mathrm{FM},6,1.73),
  (\mathrm{SDK},1,5.80),
  (\mathrm{PM},1,973.39),
  (\mathrm{TMK},2,17.58)
 \})\}}_{=\textcolor{red}{A_Q}}
). \\
\multicolumn{1}{c}{} \\
\hline
\multicolumn{1}{c}{} \\
%%%%%%%%%%%%%%%%%
%%% PROV DECK %%%
%%%%%%%%%%%%%%%%%
\multicolumn{1}{l}{\text{\textbf{ProvDeck:}}} \\
D_1 & =
(
\text{deck}_1,
\{C_{\text{agent}_1}\},
\{C_{\text{dataset}_1}, C_{\text{metadata}_1}, C_{\text{dataset}_2}\},
\{C_{\text{activity}_1}\}) \\  
\multicolumn{1}{c}{} \\
\hline
\multicolumn{1}{c}{} \\
%%%%%%%%%%%%%%%%%%
%%% PROV GRAPH %%%
%%%%%%%%%%%%%%%%%%
\multicolumn{1}{l}{\text{\textbf{ProvGraph:}}} \\
G & = (V,E) \\
V & = \{
(\text{dataset}_1,\text{entity},A_1),
(\text{metadata}_1,\text{entity},A_2),
(\text{activity}_1,\text{activity},A_3),
(\text{agent}_1,\text{agent},A_4),
(\text{dataset}_2,\text{entity},A_5)
\} \\
E & = \{
(\text{dataset}_1,\text{activity}_1,\texttt{used}),(\text{metadata}_1,\text{activity}_1,\texttt{used}),(\text{agent}_1,\text{activity}_1,\texttt{wasAssociatedWith}),
\\
& \quad\ (\text{activity}_1,\text{dataset}_2,\texttt{wasGeneratedBy}),
(\text{dataset}_1,\text{dataset}_2,\texttt{wasDerivedFrom})
\}.
\end{array}
$}
\end{table*}

\begin{definition}[ProvDeck]
Let $id_D$ be a unique identifier and $C_D\in\mathcal{P}_{\mathrm{fin}}(\mathcal{C})$ a finite set of ProvCards.
Then, a \textit{ProvDeck} is defined as the tuple $D=(id_D,C_D\!\mid_{\mathrm{agent}},C_D\!\mid_{\mathrm{entity}},C_D\!\mid_{\mathrm{activity}})$.
\end{definition}

The separation between ProvCards and ProvDecks is central to the artifact-oriented design. ProvCards can be generated and exchanged independently, whereas ProvDecks define which artifacts are considered together for a particular provenance context or exchange operation. A ProvDeck therefore establishes provenance scope without changing the semantics or identity of the contained artifacts.

%%%%%%%%%%%%%%%%%%%%%%%%%%%%%%%%%%%%%%%%%%%%%%%%%%%%%%%%%%%%%%
\subsection{Provenance Graphs}
\label{sec:4-provModel:ProvGraph}

\textit{ProvGraphs} provide a derived analytical representation of the provenance artifacts contained in a ProvDeck. They enable provenance queries, visualization, dependency analysis, and graph-based exploration without requiring the original ProvCards to be merged into a single artifact.
ProvGraphs follow the graph semantics of PROV-DM~\cite{provDM}, where nodes represent entities, activities, and agents and edges represent provenance relations. Thus, a ProvGraph is not an additional source of provenance information, but a derived view that makes the relationships encoded by a collection of ProvCards explicit for analysis.

\begin{definition}[ProvGraph]
Let $C_D \subseteq \mathcal{C}$ be the finite set of ProvCards contained in a ProvDeck $D$.
Let $V=\{v_1,\ldots,v_n\}$ be the set of nodes extracted from ProvCards, where each node $v_i$ corresponds to exactly one ProvCard $C_i \in C_D$. The following holds:
$v_i=(id_{C_i},t_{C_i},A_{C_i})$, %$v_i=(id_{C_i},t_{C_i},A_{C_i},A_{Q_i})$,
with $id_{C_i}$ a unique identifier, $t_{C_i}\in\mathcal{T}$ a PROV type,
$A_{C_i}$ a set of provenance attributes.
Quality aspects $A_{Q_i}$ associated with a ProvCard are not part of the ProvGraph representation.

Let $E=\{e_1,\ldots,e_m\}$ be the set of provenance edges extracted from the relationships contained in ProvCards. Each edge is defined as
$e=(id_{C_S},id_{C_T},r)$,
with $id_{C_S}$ and $id_{C_T}$ denoting source and target identifiers and $r\in\mathcal{E}$.
Then, a \textit{ProvGraph} is defined as the directed attributed graph $G=(V,E)$.
\end{definition}

The graph representation intentionally provides a projection of the information represented by the ProvCards. It preserves the identifiers, PROV types, provenance attributes, and provenance relationships required for graph-based analysis, while quality aspects and artifact-specific information are omitted.

%%%%%%%%%%%%%%%%%%%%%%%%%%%%%%%%%%%%%%%%%%%%%%%%%%%%%%%%%%%%%%
\newpage
\subsection{Provenance Operations and Transformations}
\label{sec:4-provModel:ProvenanceTramsformation}

We define four transformations for packing, reconstructing, and projecting provenance artifacts and characterize their information-preservation properties. These transformations are not necessarily invertible: packing may omit, abstract, or anonymize information, while graph extraction may exclude artifact-specific information. These preservation considerations also apply to quality information, which extends quality ProvCards without changing their underlying provenance semantics. Quality information may therefore be preserved, omitted, or anonymized during transformation; however, unpacking must reconstruct all quality information represented in a ProvDeck.

We first establish the following notation:
\begin{itemize}
\item $\mathcal{C}$ denotes the set of all syntactically valid (quality) ProvCards;
\item $\mathcal{P}_\mathrm{fin}(\mathcal{C})$ denotes the set of all finite subsets of $\mathcal{C}$;
\item $\mathcal{E}$ denotes the set of supported PROV relation types, including \texttt{wasGeneratedBy}, \texttt{used}, \texttt{wasDerivedFrom}, and \texttt{wasAsso-} \texttt{ciatedWith};
\item $\mathcal{D}$ denotes the set of all valid ProvDecks;
\item $\mathcal{G}$ denotes the set of all valid ProvGraphs.
\end{itemize}

\begin{figure}
\centering
\includegraphics[width=\linewidth]{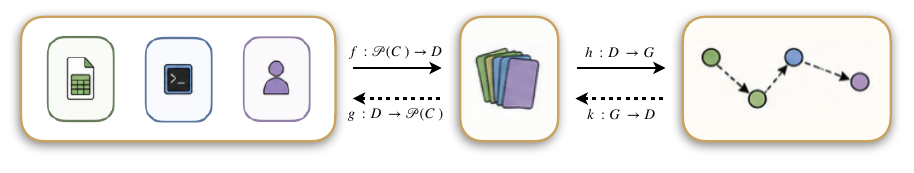}
\caption{Transformation between ProvCards, ProvDecks, and ProvGraphs within the artifact-oriented provenance model.}
\label{fig:mappings}
\end{figure}

\paragraph{\textbf{Packing and Unpacking.}}
For a finite set of (quality) ProvCards $S=\{C_1,\ldots,C_m\}\in\mathcal{P}_{\mathrm{fin}}(\mathcal{C})$, we define a \textit{packing operation} parameterized by a privacy policy $\pi$ as $f_\pi:\mathcal{P}_{\mathrm{fin}}(\mathcal{C})\rightarrow\mathcal{D}$. For a given input set $S$, the packing operation produces a ProvDeck $f_\pi(S)=D=(id_D,C_D\!\mid_{\mathrm{agent}},C_D\!\mid_{\mathrm{entity}},C_D\!\mid_{\mathrm{activity}})$, where $id_D$ is a unique identifier and $C_D$ is the set of ProvCards represented in the ProvDeck. Depending on $\pi$, the represented cards may correspond directly to cards in $S$ or to representations of cards in $S$ in which information has been omitted, abstracted, or anonymized. Any deck-level metadata or organizational information is represented by the corresponding ProvCards and their provenance relationships rather than by introducing additional elements into the ProvDeck tuple.
By default, packing preserves all provenance information contained in $S$. Depending on $\pi$, \textit{privacy-aware packing} may omit individual ProvCards or attributes, or replace attribute values by anonymized representations. Thus, information reduction is not an inherent property of packing but results from the applied policy. This separation is important for interoperability: the same provenance artifacts can be exchanged losslessly when no information-reducing policy is applied, while policy-controlled representations can be created when the exchange context imposes privacy or disclosure constraints.

Let $\operatorname{Info}(X)$ denote the provenance and quality information explicitly represented by $X$, where $X$ is a finite set of ProvCards, a ProvDeck, or a ProvGraph. These information includes identifiers, PROV types and subtypes, attributes, quality aspects, and provenance relationships.
ProvDeck identifiers and representation-specific ordering or serialization details are excluded because they organize the representation rather than describe provenance. \textit{Information preservation} between two representations $X$ and $Y$ is expressed by $\operatorname{Info}(X)\subseteq\operatorname{Info}(Y)$. Thus, all information represented by $X$ is preserved in $Y$, while $Y$ may contain additional information.
For a packing policy $\pi$ that only omits ProvCards or attributes, the packing operation satisfies $\operatorname{Info}(f_\pi(S)) \subseteq \operatorname{Info}(S)$.
If $\pi$ preserves all information contained in $S$, packing is lossless and $\operatorname{Info}(f_\pi(S)) = \operatorname{Info}(S)$.
Policies that replace values through abstraction or anonymization produce representations derived from $S$, but the explicitly represented information items do not necessarily form a literal subset of $\operatorname{Info}(S)$.

Conversely, we define the \textit{unpacking operation} $g:\mathcal{D}\rightarrow\mathcal{P}_{\mathrm{fin}}(\mathcal{C})$ by $g(D)=S_D'$, where $S_D'$ is the finite set of ProvCards reconstructed from the information represented in $D$. Unpacking is required to reconstruct all information represented in the ProvDeck without introducing additional provenance information. Formally, this requirement is expressed as $\operatorname{Info}(g(D))=\operatorname{Info}(D)$. Thus, information intentionally omitted or abstracted during packing cannot in general be recovered during unpacking. If information has been anonymized, the corresponding anonymized representation must be preserved during unpacking, without implying recoverability of the original value.

Accordingly, $g$ is a reconstruction counterpart to $f_\pi$ rather than a classical inverse. In general, $g(f_\pi(S))\neq S$, because packing may intentionally omit, abstract, or anonymize information. Since unpacking reconstructs exactly the information represented in the ProvDeck, we have $\operatorname{Info}(g(f_\pi(S)))=\operatorname{Info}(f_\pi(S))\subseteq\operatorname{Info}(S)$. If $\pi$ preserves all information, the reconstruction is lossless, including the quality information represented by the packed ProvCards. 

\paragraph{\textbf{Graph Extraction.}} 
The second transformation maps a ProvDeck to a graph representation. We define the
\textit{graph extraction operation} $h:\mathcal{D}\rightarrow\mathcal{G}$.
For a given ProvDeck $D=(id_D,C_D\!\mid_{\mathrm{agent}},C_D\!\mid_{\mathrm{entity}},C_D\!\mid_{\mathrm{activity}})$, the extracted ProvGraph is defined as $h(D)=G=(V,E)$. Its node set $V$ is $V=\{(id_{C_i},t_{C_i},A_{C_i})\mid C_i\in C_D\}$. Thus, each ProvCard contributes exactly one graph node whose identifier, PROV type, and provenance attributes are preserved.
The edge set $E$ is constructed from the provenance relationships represented by the ProvCards. For each pair of ProvCards $C_i,C_j\in C_D$, an edge $(id_{C_i},id_{C_j},e)$ is included in $E$ if $(id_{C_j},t_{C_j},e)\in R_{\mathrm{out}}^i$ or $(id_{C_i},t_{C_i},e)\in R_{\mathrm{in}}^j$, where $e\in\mathcal{E}$. Thus, the graph captures the provenance relationships whose source and target cards are contained in the ProvDeck. 
Quality aspects are stored as dedicated nodes in the Artifact Database (cf.\,Section~\ref{sec:4-provModel-inforstructure}) but are not included in the ProvGraph produced by $h$. Consequently, graph extraction preserves the core provenance structure while omitting $A_Q$.

\paragraph{\textbf{Graph Reconstruction.}}
Conversely, a ProvGraph can be used to reconstruct a ProvDeck. We therefore define the \textit{graph reconstruction operation} $k:\mathcal{G}\rightarrow\mathcal{D}$. For a given ProvGraph $G=(V,E)$, where each node is defined as $v_i=(id_i,t_i,A_i)$, we first reconstruct a set of ProvCards $C_G=\{C_1,\ldots,C_n\}$.
For each graph node $v_i$, the corresponding ProvCard is reconstructed as $C_i=(id_i,t_i,t_{S_i},A_i,R_{\mathrm{in}}^i,R_{\mathrm{out}}^i)$, where $t_{S_i}$ is assigned according to the applicable ProvCard specification. The incoming and outgoing relationships are reconstructed from the graph edges: $R_{\mathrm{out}}^i = \{(id_j,t_j,e)\mid(id_i,id_j,e)\in E\}$ and $R_{\mathrm{in}}^i = \{(id_j,t_j,e)\mid(id_j,id_i,e)\in E\}$.

Since quality aspects $A_Q$ are not represented in the ProvGraph, they cannot be recovered by the graph reconstruction operation $k$. A ProvCard reconstructed solely from $G$ therefore contains the core provenance information represented by its graph node and edges, but not the quality information stored in the corresponding quality ProvCard. Recovering this information requires access to the corresponding ProvCards, ProvDeck, or Artifact Database.
The reconstructed ProvCards are classified according to their PROV types
and assembled into a new ProvDeck $k(G)=D_G= (id_{D_G}, C_G\!\mid_{\mathrm{agent}}, C_G\!\mid_{\mathrm{entity}}, C_G\!\mid_{\mathrm{activity}})$, where $id_{D_G}$ is a newly assigned unique identifier. The newly assigned ProvDeck identifier has no effect on the reconstructed provenance information and is therefore excluded from comparisons based on $\operatorname{Info}$.

Graph reconstruction is therefore not a strict inverse of graph extraction. It preserves the core provenance information represented by the input graph, such that $\operatorname{Info}(G)\subseteq\operatorname{Info}(k(G))$. The reconstructed ProvDeck may additionally contain information required by the applicable ProvCard specification, such as assigned subtypes. Its newly assigned deck identifier is representation-specific and is not considered part of the reconstructed provenance information. However, quality aspects and other artifact-specific information omitted during graph extraction cannot be recovered from the ProvGraph alone. Consequently, $k(G)$ is semantically consistent with $G$ but need not be identical to the ProvDeck from which $G$ was extracted.

\paragraph{\textbf{Information Preservation.}}
The transformations can therefore be characterized by the information
relation. For packing and unpacking, we obtain $\operatorname{Info}(g(f_\pi(S))) = \operatorname{Info}(f_\pi(S)) \subseteq \operatorname{Info}(S)$. For graph extraction and reconstruction, we obtain $\operatorname{Info}(h(D)) \subseteq \operatorname{Info}(D)$ and $\operatorname{Info}(h(D)) \subseteq \operatorname{Info}(k(h(D)))$. The latter relation expresses that graph reconstruction preserves the information represented by the graph, but does not imply recovery of information that was not represented in the graph projection.

\paragraph{\textbf{Provenance Lifecycle.}}
Together, these transformations support a provenance lifecycle in which different components may perform capture, composition, analysis, and preservation. Workflow components generate ProvCards during or after execution without requiring a global workflow representation. Related ProvCards can subsequently be aggregated into ProvDecks and transformed into ProvGraphs for querying, visualization, and dependency analysis. Because ProvCards and ProvDecks can be stored independently of the systems that generated them, the provenance remains available for later exchange, reuse, and analysis.

%%%%%%%%%%%%%%%%%%%%%%%%%%%%%%%%%%%%%%%%%%%%%%%%%%%%%%%%%%%%%%
\subsection{Provenance Infrastructure}
\label{sec:4-provModel-inforstructure}

Operationalizing this lifecycle requires infrastructure for generating, persisting, retrieving, and composing provenance artifacts.

\paragraph{\textbf{ProvTemplate.}} 
When the structure of a preprocessing pipeline is known in advance, a \textit{ProvTemplate} can be used to define the expected provenance structure. It consists of a predefined set of ProvCards, grouped within a ProvDeck, representing the expected workflow components and their relationships. During execution, the template is instantiated by removing unused ProvCards and relationships and  populating the remaining cards with concrete provenance information. This reduces the effort required to collect and organize provenance metadata during runtime.

\paragraph{\textbf{Artifact Database.}}
The artifact-oriented model can be complemented by an optional persistent repository for collecting, indexing, preserving, and reusing ProvCards. The database is not required for provenance capture or exchange, but supports discovery, retrieval, versioning, long-term preservation, provenance analysis, workflow comparison, and reuse across heterogeneous workflow environments.
\section{Implementation}
\label{sec:5-implementation}

\begin{figure}
\centering
\includegraphics[width=0.95\linewidth]{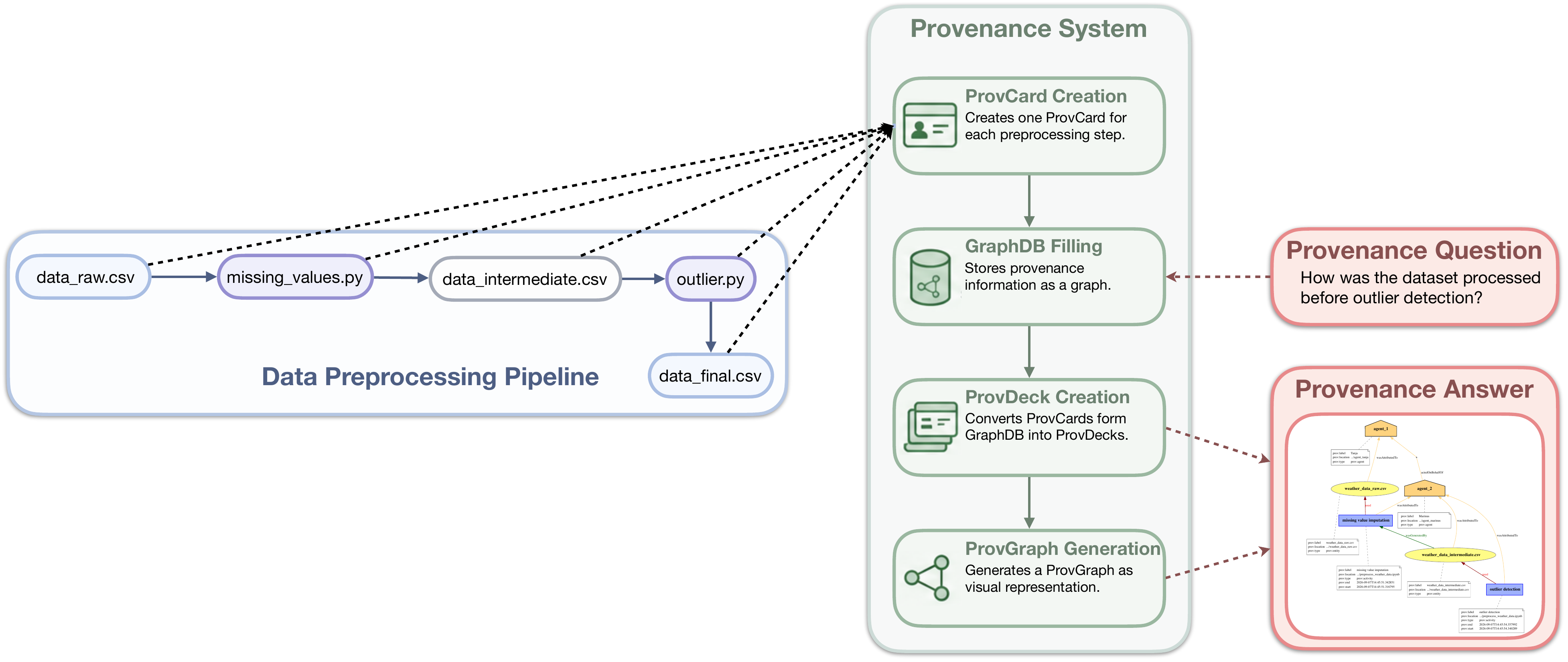}
\caption{Software architecture with application-specific input (blue), provenance system (green) and provenance questions answering (red).}
\label{fig:implementation}
\end{figure}

To operationalize the proposed model and infrastructure, our implementation
comprises four main steps (cf.\,Figure~\ref{fig:implementation}):
(1) ProvCard creation, 
(2) storage in a graph database,
(3) ProvDeck creation, and
(4) ProvGraph generation.
These steps involve three software components: a database component for storing and retrieving ProvCards, a transformation component for converting between ProvCards, ProvDecks, and W3C PROV-JSON, and a visualization component for generating ProvGraphs.

\paragraph{\textbf{ProvCard Creation.}}
Provenance extraction and ProvCard creation constitute the application-specific part of the implementation. We distinguish between two alternative integration strategies: 
In the first strategy, the executing component collects the relevant information, including input and output artifacts, responsible agents, execution timestamps, and optional quality information. This information is immediately transformed into ProvCards and stored in the Artifact Database. Consequently, provenance capture is interleaved with data processing, and the corresponding provenance representation is available as soon as the workflow step has completed.

In the second strategy, provenance information is reconstructed retrospectively from execution logs. During workflow execution, the participating components write provenance-relevant events, such as accessed inputs, generated outputs, timestamps, parameters, and responsible processes, to their respective log files. After the workflow has completed, workflow-specific extractors parse these heterogeneous log files, map the extracted information to the ProvCard model, and import the resulting ProvCards into the Artifact Database.
The two strategies differ in the time and location of provenance extraction but produce the same standardized representation. Direct capture requires explicit instrumentation of the executing workflow but avoids a separate ingestion phase and makes provenance available immediately. Retrospective extraction introduces less provenance-specific logic into the workflow itself, but its completeness depends on the information recorded in the available logs and requires workflow-specific parsing and mapping. 

\paragraph{\textbf{Provenance Model.}}
Once created, the ProvCards provide a standardized representation of the extracted provenance information based on our provenance model. All subsequent components are implemented independently of the underlying use case and workflow system. This separation establishes a clear boundary between application-specific provenance extraction and application-indepen\-dent provenance processing. Consequently, the same processing pipeline can be applied to different workflow systems without modifying the use-case-independent components.

\paragraph{\textbf{ProvDeck Creation.}}
A ProvDeck comprises a set of related ProvCards and can be serialized as a JSON file. Each ProvCard is represented by a JSON object, as illustrated in Listing~\ref{lst:ProvCard}. For quality ProvCards, \(A_Q\) is represented by the quality object. Its property names correspond to quality-aspect identifiers \(q_i\), while the associated property values represent the structured values \(w_i\). While the JSON representation provides a portable exchange format for complete ProvDecks, the individual ProvCards and their relationships are persistently stored in the Neo4j database.

\begin{figure}[t]
\centering
\begin{lstlisting}[language=json,firstnumber=1, caption=JSON serialization of a quality ProvCard for an adaptive missing-value imputation activity., label=lst:ProvCard]
{
  "id": "prov::activity::missing_value_imputation_01",
  "type": "activity:code",
  "name": "Adaptive missing-value imputation",
  "location": "workflow://missing_value_imputation_01",
  "quality": {
    "imputation_gap_37": {
      "primary_recommendation": "seasonal",
      "applied_method": "linear",
      "decision_status": "fallback"
    }
  },
  "input": [
    {
      "id": "prov::data::input_series",
      "type": "data",
      "edge": "used"
    }
  ]
}
\end{lstlisting}
\end{figure}

\paragraph{\textbf{Artifact Database.}}
The artifact database stores provenance artifacts in a graph-oriented representation. 
This representation supports efficient traversal of provenance paths, dependency analysis, and graph queries while preserving the semantics of the underlying W3C PROV-DM model.
Each ProvCard (describing an entity, activity, or agent) is mapped to a graph node.
The corresponding PROV type determines the node label, while the metadata attributes in the attribute set $A$ are represented as node properties. 
The relationships specified in $R_\text{in}$ and $R_\text{out}$ are materialized as directed graph edges representing PROV relations such as \texttt{used}, \texttt{wasGeneratedBy}, \texttt{wasDerivedFrom}, and \texttt{wasAssociatedWith}.

\paragraph{\textbf{Quality Information.}} 
Quality information represented by quality ProvCards is stored separately from the corresponding provenance nodes. While $A_Q$ is conceptually associated with the ProvGraph node representing the ProvCard, the physical Neo4j representation stores each structured quality aspect in a dedicated node to facilitate querying and aggregation. Each quality aspect $(q_i,w_i) \in A_Q$ is connected to its associated provenance node through a \texttt{has\_quality} relationship. The identifier $q_i$ identifies the quality aspect, while the key-value pairs contained in $w_i$ are represented as properties of the quality node. 

\paragraph{\textbf{ProvGraph Generation.}}
The final implementation step generates a graphical representation of the provenance information. For this, a W3C PROV-JSON file is generated and converted into a directed graph using the \texttt{prov} library. To improve readability, entity labels containing file paths are reduced to their corresponding file names before rendering. 

\section{Case Study}
\label{sec:6-study}

\begin{figure*}[t]
\centering
\begin{subfigure}[b]{0.35\textwidth}
\centering
\includegraphics[width=\textwidth]{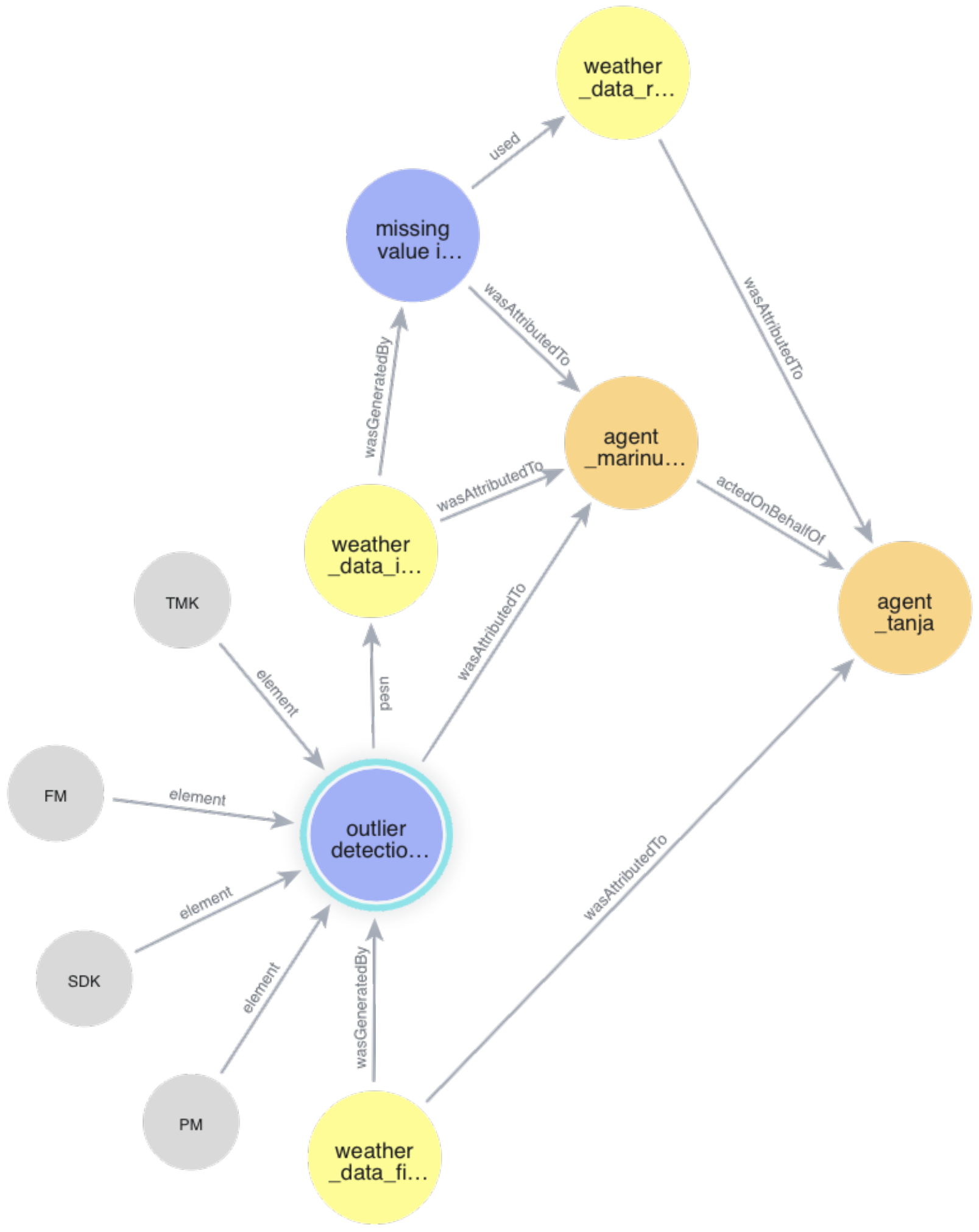}
\caption{Artifact Database representation including all ProvCards.}
\label{fig:graphDB}
\end{subfigure}
\hfill
\begin{subfigure}[b]{0.625\textwidth}
\centering
\includegraphics[width=.8\textwidth]{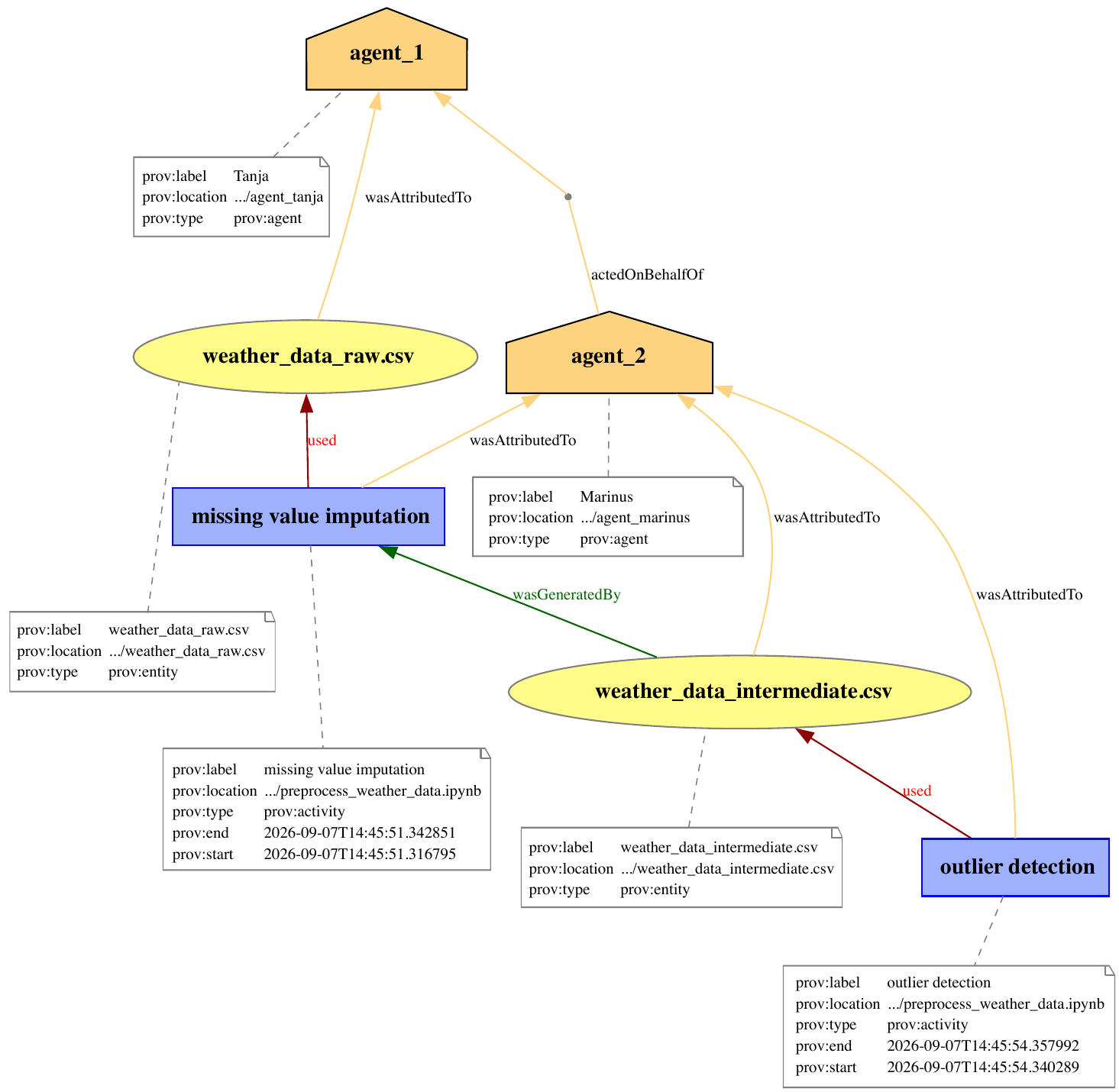}
\caption{Query-specific ProvGraph answering provenance question $Q$}
\label{fig:provGraph}
\end{subfigure}
\caption{Graph-based answering of the provenance question \textit{Q: How was the dataset processed before outlier detection?} (a) Complete graph representation of all ProvCards stored in the Artifact Database for the case-study workflow, including quality information. (b) Query-specific ProvGraph containing only the provenance elements required to answer the question.}
\label{fig:caseStudy}
\end{figure*}

To evaluate the proposed provenance framework, we consider a data-processing workflow that performs two consecutive preprocessing steps on a weather dataset. The workflow (cf.\,Figure~\ref{fig:workflow}) is deliberately kept simple to illustrate how provenance information is captured, stored, and queried while still covering different types of provenance questions.

\paragraph{\textbf{Dataset.}}
The workflow processes daily wind speed, sunshine duration, air pressure, and temperature. 
Missing values, encoded as $-999$, are imputed by \texttt{missing\_values.py}, after which \texttt{outlier.py} identifies implausible measurements. The resulting raw, intermediate, and final datasets form the entities of the provenance path shown in Figure~\ref{fig:workflow}.

\paragraph{\textbf{Provenance Capture.}}
The implementation evaluated in this paper applies the direct-capture strategy described in Section~\ref{sec:5-implementation}. The preprocessing workflow and its provenance instrumentation are implemented in a single Jupyter notebook. After each processing step, a dedicated cell creates the corresponding entity, activity, and agent ProvCards, records their PROV relationships, timestamps, and optional quality information, and stores them in the Neo4j-based Artifact Database. ProvCards are thus generated incrementally during workflow execution. Retrospective extraction from log files is supported as an alternative strategy but is not implemented or evaluated in this case study.

\paragraph{\textbf{ProvDeck Creation.}} 
The ProvCards generated throughout the preprocessing workflow are collected into a complete ProvDeck, preserving their individual identities and relationships. The size of a ProvDeck depends on the number and detail of its ProvCards. In this case study, the serialized complete ProvDeck has a size of 6 kB.

\paragraph{\textbf{Provenance Querying.}}
Provenance questions are translated into Cypher queries and executed against the Artifact Database. The graph-oriented representation allows the queries to traverse provenance relations between entities, activities, and agents. Depending on the query, the result can consist of individual provenance attributes, a set of provenance elements, or a connected subgraph of the stored provenance graph.

The resulting subgraph is reconstructed as a ProvDeck, preserving the individual provenance artifacts and their relationships, and represented as a ProvGraph for graph-based analysis and visualization. For queries that only require a textual result, the query result can alternatively be returned as a string. This allows the same provenance representation to support both machine-readable provenance artifacts and human-readable query results.
As an example of a graph-based provenance query, we consider the question \textit{Q: How was the dataset processed before outlier detection?} Answering this question requires following the provenance relations from the outlier detection activity to its input artifacts and further back to the preceding processing activity and its inputs.

The corresponding Cypher query retrieves the provenance elements and relations required to reconstruct the processing history preceding the outlier-detection activity. The resulting query-specific ProvGraph is shown in Figure~\ref{fig:provGraph}. It shows that \texttt{weather\_} \texttt{data\_raw.csv} is used by \texttt{missing\_values.py}, which generates \texttt{weather\_} \texttt{data\_intermediate.csv}. This intermediate dataset subsequently serves as the input to \texttt{outlier.py}. Because the question concerns processing before outlier detection, the output of that activity, \texttt{weather\_data\_final.csv}, is not included in the query result. In contrast, Figure~\ref{fig:graphDB} shows the complete Artifact Database representation, including quality information and the final dataset \texttt{weather\_} \texttt{data\_final.csv}.

\section{Validation of the Provenance Model}
\label{sec:7-validation}

We validate three aspects of the proposed model: requirement coverage, the information-preservation properties defined in Section~\ref{sec:4-provModel}, and the reconstruction and querying of independently captured provenance. Performance, scalability, capture overhead, and cross-system portability are outside the scope of this evaluation.

\newpage
\paragraph{\textbf{Requirement Coverage.}}

\begin{table*}[t]
\centering
\caption{Requirement coverage and evaluation evidence.}
\label{tab:req-eval}
\begin{tabular}{p{0.5cm}p{1.85cm}p{13.5cm}}
\toprule
\textbf{Req.} & \textbf{Addressed by} & \textbf{Evidence} \\
\midrule
R1 &
ProvCards &
Structured representations capture PROV elements, relationships, and optional quality information. \\
R2 &
ProvCards &
Direct capture was integrated into the existing notebook workflow without replacing its infrastructure. \\
R3 &
ProvCards &
Workflow steps generate ProvCards independently of a global provenance representation. \\
R4 &
ProvCards, ProvDecks &
Datasets, metadata, activities, and agents are represented through a common abstraction. \\
R5 &
ProvCards, ProvDecks &
ProvCards are uniquely identified, stored in Neo4j, retrieved, and collected into ProvDecks. \\
R6 &
ProvDecks &
Independently serialized, mutually compatible cards are collected into a common ProvDeck while preserving their identities and relationships. \\
R7 &
ProvGraphs &
ProvDecks are transformed into ProvGraphs that reconstruct the expected processing path and support graph-based queries. \\
\bottomrule
\end{tabular}
\end{table*}

Table~\ref{tab:req-eval} summarizes how the implementation and case study address requirements \ref{item:R1}--\ref{item:R7}.

\paragraph{\textbf{Transformation Validation.}}
We validated the implemented transformations using the ProvCards generated for the weather-data preprocessing workflow. First, we packed the cards for the raw, intermediate, and final datasets, the missing-value imputation and outlier-detection activities, and the associated agents into a ProvDeck using the identity policy. After unpacking the deck, we compared the reconstructed cards with the original cards in terms of identifiers, PROV types and subtypes, attributes, relationships, quality-aspect identifiers, and quality values. After ordering semantically unordered collections by their identifiers, keys, and relation types, no differences were observed.

We then transformed the ProvDeck into a ProvGraph and reconstructed a ProvDeck from that graph. The graph preserved the provenance path from \texttt{weather\_data\_raw.csv} through missing-value imputation and \texttt{weather\_data\_intermediate.csv} to outlier detection and \texttt{weather\_data\_final.csv}, together with its graph-relevant attributes and relationships. Comparing these elements with the reconstructed ProvDeck revealed no differences. Quality aspects and other information excluded by the graph projection, as well as the newly assigned ProvDeck identifier, were excluded from this comparison. The results are consistent with the information-preservation properties defined in Section~\ref{sec:4-provModel:ProvenanceTramsformation}; they do not establish those properties for arbitrary implementations or information-reducing policies.

\newpage
\paragraph{\textbf{Independent Capture and Provenance Questions.}}
The ProvCards generated at different stages of workflow execution were successfully connected through compatible identifiers and collected into a ProvDeck without maintaining a shared in-memory ProvGraph during preprocessing. 
The complete ProvGraph reconstructed from these cards contained the expected processing path from \texttt{weather\_} \texttt{data\_raw.csv} through missing-value imputation and \texttt{weather\_} \texttt{data\_intermediate.csv} to outlier detection and \texttt{weath-} \texttt{er\_data\_final.csv}. The query-specific ProvGraph in Figure~\ref{fig:provGraph} contained only the prefix required to answer $Q$ and therefore excluded the final dataset. 

Following the provenance-question classification of \cite{WHA25}, we evaluated 14 questions covering \emph{how}, \emph{where}, \emph{who}, \emph{what}, \emph{which}, \emph{when}, and \emph{why}. For example, one question asks how the dataset was processed before outlier detection (Figure~\ref{fig:provGraph}). The complete set of questions and queries is available in the Git repository (see below). Eleven questions were answered entirely from the recorded provenance attributes, quality information, and relationships; three required supplementary context from the notebook, filesystem, or execution environment. All 14 answers were consistent with the known workflow and its recorded execution context. These results show that independently generated ProvCards support workflow reconstruction and provenance querying, while the range of answerable questions depends on the contextual information captured.

\paragraph{\textbf{Discussion.}}
The validation is limited to direct provenance capture in a small Jupyter-based workflow and a single execution environment. Although the architecture supports retrospective generation of ProvCards from logs, its completeness depends on the available log information and workflow-specific extraction and mapping. Evaluating this strategy, as well as capture overhead, scalability, and cross-system portability, remains future work.
\section{Conclusion and Future Work}
\label{sec:8-conclusion}

This paper introduced ProvCards, an artifact-oriented approach for capturing and composing provenance across heterogeneous scientific workflows. ProvCards represent independently manageable provenance artifacts grounded in PROV-DM, ProvDecks provide scopes for their composition and exchange, and ProvGraphs provide derived representations for querying and analysis. The formalized transformations make explicit which provenance and quality information is preserved or omitted, including information reduction through privacy-aware packing and graph projection.

Our implementation and case study demonstrate that provenance can be composed through compatible identifiers, and queried as a connected provenance graph. The results provide initial evidence for the feasibility of the approach, but are limited to a small workflow and direct capture within a single environment.

Future work will evaluate capture overhead, storage requirements, query performance, and interoperability in larger distributed workflows and across independently administered systems. We also plan to investigate reusable extraction adapters, privacy-aware provenance exchange, controlled disclosure, and extensions of the Artifact Database for discovery, versioning, workflow comparison, and long-term preservation.

\paragraph{\textbf{Code and Data Availability.}}
The source code, executable Jupyter notebook, Cypher queries, generated ProvDecks, and ProvGraphs used in this evaluation are available at \url{https://gitlab.com/TanjaAuge/provcardmodel}. The underlying weather data originate from the German Weather Service and are available at \url{https://opendata.dwd.de}.

\bibliographystyle{ACM-Reference-Format}
\bibliography{references}

\appendix

\end{document}